\documentclass[twocolumn,superscriptaddress]{revtex4-1}
\usepackage{graphicx}
\usepackage{lineno}
\usepackage{color}
\begin{document}

\title{Ir \textit{d}-band Derived Superconductivity in the Lanthanum-Iridium System LaIr$_{3}$}

\author{Neel~Haldolaarachchige}
\address{Department~of~Physics~and~Astronomy,~Rowan~University,~Glassboro,~NJ08028,~USA~}
\address{Department~of~Physics~and~Engineering,~College~of~Staten~Island,~The City University of New York,~NY10314,~USA~}
\author{Leslie~Schoop}
\address{Department~of~Chemistry,~Princeton~University,~Princeton,~NJ08540, USA}
\author{Mojammel A.~Khan}
\address{Department~of~Physics~and~Astronomy,~Louisiana~State~University,~Baton~Rouge,~LA70803,~USA~}
\author{Wenxuan~Huang}
\address{Department~of~Mat.~Sci.~and~Eng.,~Massachusetts~Institute~of~Technology,~Cambridge, MA 02142,~USA}
\author{Huiwen~Ji}
\address{Department~of~Chemistry,~Princeton~University,~Princeton,~NJ08540, USA}
\author{Kalani~Hettiarachchilage}
\address{Department~of~Physics~and~Engineering,~College~of~Staten~Island,~The City University of New York,~NY10314,~USA~}
\address{Department~of~Physics,~The~College~of~New~Jersey,~Ewing,~NJ08618,~USA~}
\author{David~P.~Young}
\address{Department~of~Physics~and~Astronomy,~Louisiana~State~University,~Baton~Rouge,~LA70803,~USA~}

\begin{abstract}  
The electronic properties of the heavy metal superconductor LaIr$_{3}$ are reported. The estimated superconducting parameters obtained from physical properties measurements indicate that LaIr$_{3}$ is a BCS-type superconductor. Electronic band structure calculations show that Ir \textit{d}-states dominate the Fermi level. A comparison of electronic band structures of LaIr$_{3}$ and LaRh$_{3}$ shows that the Ir-compound has a strong spin-orbit-coupling effect, which creates a complex Fermi surface. 
\end{abstract}

\maketitle
\newcommand{\angstrom}{\mbox{\normalfont\AA}}

\section{Introduction}
The combination of rare earth and main group transition metal elements can produce compounds with complex, technologically important, and scientifically interesting physical properties. Compounds with 5\textit{d} transition metals show electron correlations and spin-orbit interactions that create interesting electronic properties, such as superconductivity.~\cite{iridate}
Only a few Ir compounds are known to display superconductivity and most of their electronic properties are due to the rare earth elements rather than Ir.~\cite{ceirin5, ca3irge32014, cairge32010, daigo2013} There are, however, a few examples, such as CaIr$_{2}$, IrGe and Mg$_{10}$Ir$_{19}$B$_{16}$, where the superconductivity is derived from Ir-\textit{d} states at the Fermi surface.~\cite{neel_cair2, daigo2014, mgirb} 
Evens so, there is no simple example of a lanthanide-iridium superconductor whose properties are dominated by Ir-\textit{d} bands at the Fermi level and strong spin-orbit-coupling. 

Here we report the synthesis, experimental superconducting parameters of LaIr$_{3}$ and LaIr$_{2.8}$ and the calculated electronic band structure of LaIr$_{3}$ and LaRh$_{3}$. 
The existence of superconductivity in LaIr$_{3}$ has been reported earlier, but besides its superconducting transition temperature T$_{c}$, no further characterization is available, except one report on a doping study.~\cite{lair3_1, lair3_2} LaIr$_{3}$ is unusual among the lanthanide superconductors, because it contains 5\textit{d} electrons with no magnetic rare earth ions present. The reported T$_c$ of LaIr$_{3}$ is, however, comparable to most Ce 4\textit{f}  based superconductors.~\cite{ ceru2-1, ceru2-2}
Still, the superconducting transition temperature of LaIr$_{3}$ is much lower than those of the alkaline-based Ir superconductors such as CaIr$_{2}$.~\cite{neel_cair2} The isostrucutral compound LaRh$_{3}$ was also reported to be superconducting with a slightly lower than that of LaIr$_{3}$.~\cite{larh3}

\section{Experiment and Calculation}
Polycrystalline samples of LaIr$_{3}$ were prepared with ultra pure (5N, 99.999\%) elemental La and Ir by arc melting. The arc-melted button was sealed inside a quartz tube under vacuum. The tube was then heated (heating rate is 180~$^{0}$C per hour) to 1000 $^{0}$C and held at that temperature for about 24 hrs. The quartz tubes with the samples inside were then removed from the furnace and quenched in water. 
Powder X-ray diffraction (PXRD) were performed to check the phase purity and cell parameters of the samples at room temperature on a Bruker D8 FOCUS diffractometer (Cu$~K_{\alpha}$). MAUD program was used for detailed Rietveld analysis.~\cite{maud} The X-ray investigation confirmed that the sample was a single phase with a rhombohedral crystal symmetry with the space group 166 (R -3 m).~\cite{lair3_struc} A small amount of elemental Ir was detected as an impurity, but this dose not effect on the observed physical properties of the sample (Ir metal superconduct at very low temperature 0.14K).~\cite{ir-sc} Energy Dispersive Spectroscopy (EDS) is used for further investigation of the sample purity by using FEI XL30 FEG-SEM system equipped with EVEX EDS, and confirmed the stoichiometry (ratio of La:Ir) of the sample as 1:3. 

Standard four-probe method was used to measure electrical resistance with an excitation current of 10 mA; small diameter Pt wires were attached to the sample using a conductive epoxy (Epotek H20E). Quantum Design Physical Property Measurement System (PPMS) was used to collect the data from 300K - 2K and in magnetic fields up to 5 T (Tesla). Time-relaxation method was used to measure specific heat between 2 K and 50 K in the PPMS, using zero field and 5 T applied magnetic field. The magnetic susceptibility was measured in a DC field of 10 Oe; the sample was cooled down to 1.8 K in zero-field, the magnetic field was then applied, and the sample magnetization was measured on heating to 4 K [zero-field-cooled (ZFC)], and then on cooling to 1.8 K [field-cooled (FC)] in the PPMS.

Density functional theory (DFT) based computational method was used to calculate the electronic structure with the WIEN2K code with a full-potential linearized augmented plane-wave and local orbitals [FP-LAPW + lo] basis~\cite{blaha2001,sign1996,madsen2001,sjosted2000} together with the PBE parameterization~\cite{perdew1996} of the GGA, with and without spin orbit coupling (SOC). The plane-wave cutoff parameter R$_{MT}$K$_{MAX}$ was set to 7, and the Brillouin zone was sampled by 20,000 k points. Convergence was confirmed by increasing both R$_{MT}$K$_{MAX}$ and the number of k points until no change in the total energy was observed. Program Crysden was used to plot the Fermi surface for the calculated band structure.

\section{Results and Discussion}

\begin{figure}[t]
  \centerline{\includegraphics[width=0.5\textwidth]{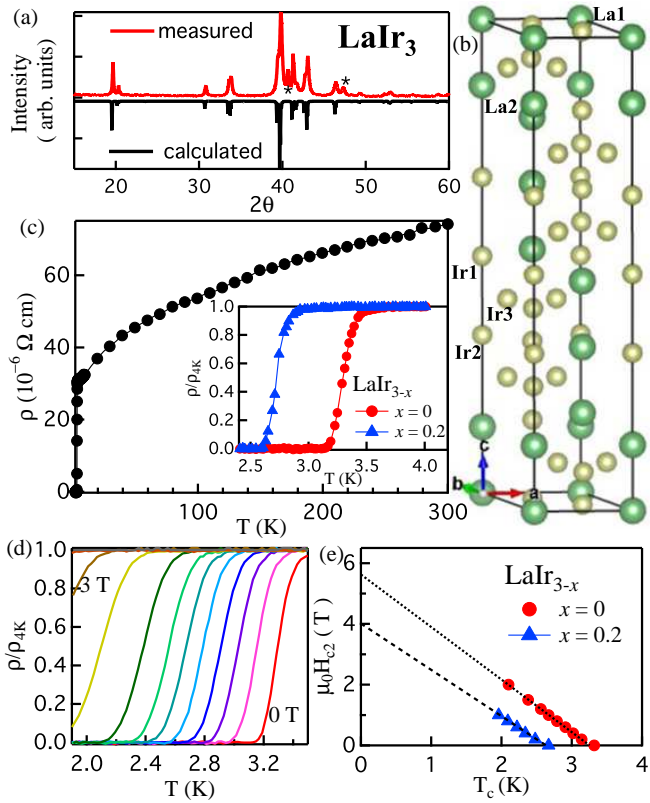}}
  \caption
    {
      (Color online)   Analysis of the powder X-ray diffraction (PXRD) and resistivity of LaIr$_{3}$. (a) Measured powder X-ray diffraction and calculated XRD patterns of LaIr$_{3}$. (b) 3D crystal structure of rhombohedral LaIr$_{3}$. (c) Resistivity at zero applied field. Inset of (c) shows normalized resistance at low temperature of LaIr$_{3}$ and LaIr$_{2.8}$. (d) Normalized resistance of LaIr$_{3}$ with increasing applied magnetic field from 0 - 3 T. (e) The upper critical field as a function of temperature. 
    }
  \label{Fig1}
\end{figure}

\begin{figure}[t]
  \centerline{\includegraphics[width=0.5\textwidth]{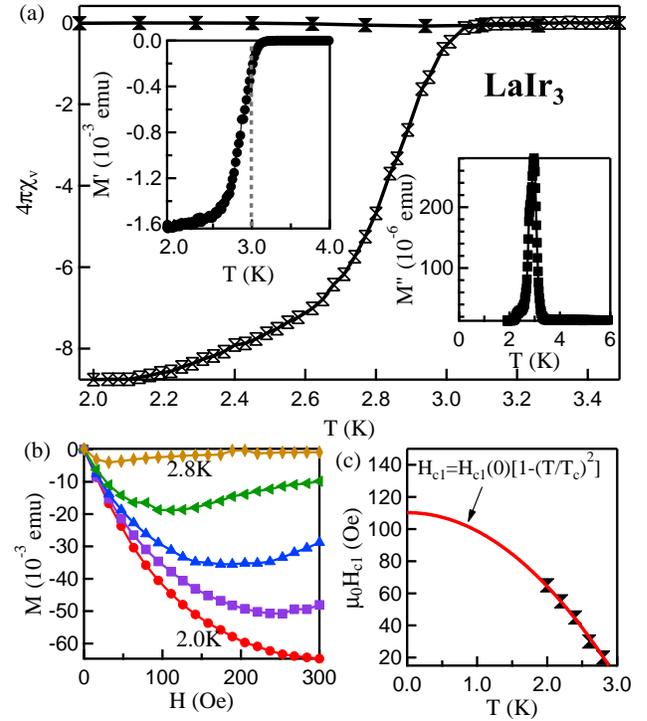}}
  \caption
    {
      (Color online)  Analysis of magnetization of LaIr$_{3}$. (a) The ZFC and FC DC magnetization of LaIr$_{3}$. Inset of (a) shows the AC magnetization. (b) DC magnetization (M vs. H curves) at different temperatures from 2K to 2.8K, below the superconducting temperature T$_{c}$. (c) The lower critical field as a function of temperature. 
    }
  \label{Fig2}
\end{figure}

Fig.~\ref{Fig1}.(a) shows measured powder X-ray diffraction (PXRD) (upper line) and calculated PXRD pattern (lower line) of polycrystalline LaIr$_{3}$. Two pattern match perfectly, which confirms the rhombohedral crystal structure with space group number 166 (R -3 m). However, very small amount of elemental Ir impurity (asterisk marks on the Fig.~\ref{Fig1}.(a)) detected on the PXRD and this does not have any effect on observed physical properties of LaIr$_{3}$ because, elemental Ir superconduct only at very low temperature (~0.14 K).~\cite{ir-sc, ir_metal} Fig.~\ref{Fig1}.(b) shows 3D crystal structure of LaIr$_{3}$. Large green symbols represents La atoms and small yellow symbols represents Ir atoms. Rhombohedral unit cell contains two distinct La atoms (La1 and La2) and three distinct Ir atoms (Ir1, Ir2 and Ir3). Fig.~\ref{Fig1}.(c) shows the temperature dependent resistivity of LaIr$_{3}$ from 300K to 2K. The normal state resistivity shows poor metallic behavior $\left( \frac{d\rho}{dT} > 0\right)$ of the polycrystalline sample, with a very low room temperature resistivity. Inset of Fig.~\ref{Fig1}.(c) shows a clear superconducting transition (T$^{onset} _{c}$) of LaIr$_{3}$ at 3.3 K and a slightly lower T$^{onset} _{c}$ for LaIr$_{2.8}$ at 2.75 K. We have tested both Ir and La off-stoichiometric effects on the superconducting T$_{c}$ of this material. La deficiency did not show a strong effect, however Ir deficiency did affect T$_{c}$ significantly. This demonstrates the strong effect of Ir bands around the Fermi level, which is consistent with the band structure calculation studies later discussed in this manuscript.
The low-temperature resistivity (above T$_{c}$, from 4 K - 25 K) data can be described by a power law $\rho = \rho _{0} + AT^{n}$ with $n=2$, which follows Fermi liquid behavior with the residual resistivity $\rho _{0} = 30~\mu \Omega cm$, and the coefficient $A = 2.35\times10^{-8}~\frac{\Omega cm}{K^{2}}$. The high temperature (above 25 K) data significantly deviates from Fermi Liquid behavior and show a tendency to saturate at higher temperatures. This behavior can be related to the Ioffe-Regel limit~\cite{ioffe}, when the charge carrier mean free path is comparable to the interatomic spacing and/or to two-band conductivity.~\cite{zverev} Similar behavior was observed for other Ir-based superconductors, such as CaIr$_{2}$.~\cite{neel_cair2}
Extracted value of \textit{A} is related to the degree of electron correlations in a system, which suggests that LaIr$_{3}$ is a weakly correlated electron system.~\cite{kittel}

Magnetoresistance data for LaIr$_{3}$ is shown in Fig.~\ref{Fig1}(d). Increasing magnetic field shows the width of the superconducting transition increases slightly. The Werthamer-Helfand-Hohenberg (WHH) expression is used to estimate the orbital upper critical field by using the 50$\%$ normal state resistivity point as the transition temperature, $\mu _{0}H_{c2}(0)=-0.693~T_{c}\frac{dH_{c2}}{dT}\vert_{T=T_{c}}$.~\cite{tinkham} Fig.~\ref{Fig1}(e) between $\mu _{0}H_{c2}$ and $T_{c}$ shows linear relationship. $\mu _{0}H_{c2}(0)=$~3.84~T is calculated by using slope of the graph and the value of $\mu _{0}H_{c2}(0)$ is smaller than the weak-coupling Pauli-paramagnetic limit $\mu _{0}H^{Pauli} = 1.82~T_{c} = 6.10$ T for LaIr$_{3}$. 
Empirical formula $H_{c2}(T)=H_{c2}(0)\left[1-\left(\frac{T}{T_{c}}\right)^{\frac{3}{2}}\right]^{\frac{3}{2}}$ is also used to calculate the orbital upper critical field and the calculated value agrees well with the value of WHH method. Both of the above formulas (WHH expression and the empirical formula) are widely used to calculate $\mu _{0}H_{c2}(0)$ for a variety of intermetallic and oxide superconductors.~\cite{amar2010, amar2011, lan2001, neel2014, maz2013, huxia2013, neel2014-2, au2bi} 
Upper critical field $\mu _{0}H_{c2}(0)$ and $\xi (0)=\sqrt{\Phi _{0}/2\pi H_{c2}(0)}=92.59$~\AA ~are used to calculate the Ginzburg-Landau coherence length, where $\Phi _{0}=\frac{hc}{2e}$ is the magnetic flux quantum.~\cite{clogston1962, werthamer1966} Calculated value is comparable to the recently reported Ir based Laves phase superconductor CaIr$_{2}$, which further suggests the effect of Ir sub-lattice on both compounds.~\cite{neel_cair2}

Fig.~\ref{Fig2} shows analysis of the magnetization of LaIr$_{3}$. Zero-field-cooled (ZFC-shielding) and field-cooled (FC-Meissner) data in Fig.~\ref{Fig2}(a) confirms the superconducting shielding fraction. The bulk superconducting transition T$^{onset} _{c}$ = 3.1 K is extracted from the graph. 
Resistivity and susceptibility measurements show similar values of T$_{c}$, which indicates the polycrystalline sample is homogeneous. Inset of Fig.~\ref{Fig2}(a) further verifies the superconducting T$_{c}$ = 3.1 K by AC susceptibility. Upper left inset shows real part of AC-susceptibility with T$^{onset} _{c}$ = 3.1 K and lower right inset shows imaginary part of AC-susceptibility with similar T$_{c}$.
The magnetization as a function of magnetic field over a range of temperatures below T$_{c}$ is shown in Fig.~\ref{Fig2}(b). To analyze the lower critical field, T$_{c}$ for each magnetic field was selected by using 2.5\% deviation from the full shielding effect. Fig.~\ref{Fig2}(c) shows $\mu_{0}H_{c1}$ as a function of T$_{c}$. $H_{c1}(T)=H_{c1}(0)\left[1-\left(\frac{T}{T_{c}}\right)^{2}\right]$ is used to analyze the lower critical field behavior and the $\mu _{0}H_{c1}$ data are well described with the above equation. $\mu _{0}H_{c1}(0)$=110.24 Oe is extracted by using least squares fit, which is much smaller than that of the cubic Laves phase CaIr$_{2}$.~\cite{neel_cair2} 

\begin{figure}[t]
  \centerline{\includegraphics[width=0.5\textwidth]{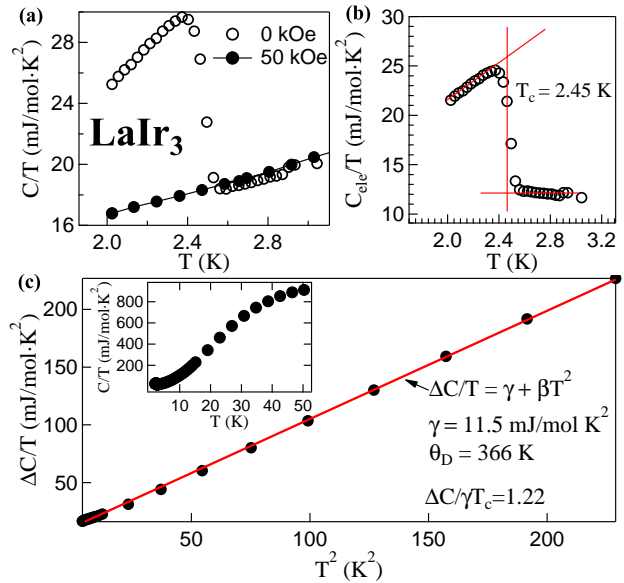}}
  \caption
    {
      (Color online)  Analysis of heat capacity of LaIr$_{3}$. (a) Heat capacity at zero (open symbol) and 5 T (Tesla) (closed symbol) applied field. (b) The electronic heat capacity jump, which can be used to extract T$_{c}$ by using the equal-area construction method. (c) The 5 T heat capacity data. The solid line shows the fit with equation $\frac{C}{T}=\gamma+\beta T^{2}$. The upper left inset shows the zero-field heat capacity as a function of temperature.    
       }
  \label{Fig3}
\end{figure}

\begin{table}[t]
\caption{Superconducting Parameters of LaIr$_{3}$.}
  \centering  
  \begin{tabular}{ lc   c   c   c }
  \hline \hline 
    Parameter & Units & LaIr$_{3}$    \\ \hline  \hline  
    $T_{c}$ & K & 3.32    \\
    $\rho _{0}$ & $m \Omega cm$ &  3.05$\times10^{-2}$     \\
    $\frac{dH_{c2}}{dT}\vert _{T=T_{c}}$ & $T~K^{-1}$ & -1.67  \\
    $\mu _{0}H_{c1}(0)$ & Oe & 110.24   \\
    $\mu _{0}H_{c2}(0)$ & T & 3.84   \\
    $\mu _{0}H^{Pauli}$ & T & 6.11   \\
    $\mu _{0}H (0)$ & T & 0.0175   \\
    $\xi (0)$ & \AA & 92.59    \\
    $\lambda (0)$ & \AA & 960     \\
    $\kappa (0)$ & \AA & 10.37  \\
    $\gamma (0)$ & $\frac{mJ}{mol~K^{2}}$ & 11.5   \\
    $\frac{\Delta C}{\gamma T_{c}}$ & & 1.22    \\
    $\Theta _{D}$ & K & 366     \\
    $\lambda _{ep}$ &  & 0.57  &     \\
    $N(E_{F})$ & $\frac{eV}{f.u.}$ & 0.72  &      \\ \hline \hline
  \end{tabular}
  \label{tab:1}
\end{table}

Fig.~\ref{Fig3} shows the analysis of specific heat (C) measurements. Specific heat jump at the thermodynamic transition is shown in Fig.~\ref{Fig3}(a). Applied magnetic field of 5T is fully suppressed the heat capacity jump. 
The superconducting transition temperature of $T_{c}$ = 2.45 K is extracted by the standard equal area construction method (see Fig.~\ref{Fig3}(b)). It is not very uncommon to see lower critical temperature from specific heat measurement compare to resistivity measurement, as heat capacity is a bulk measurement, and resistivity and susceptibility can be sensitive to surface super currents that appear at higher temperature.
The low temperature normal state specific heat is well fit with the Debye model. 
The solid line in Fig.~\ref{Fig3}(c) shows the fitting; the electronic specific heat coefficient $\gamma = 11.5 \frac{mJ}{mol~K^{2}}$, and the phonon/lattice contribution $\beta = 0.906 \frac{mJ}{mol~K^{4}}$ is extracted from the fit. 
The value of $\gamma$ obtained is comparable to the cubic Laves phase superconductor CaIr$_{2}$,~\cite{neel_cair2} and some other Ir-based heavy element superconductors, which further suggests the common effect of Ir-sub-lattice on many Ir based superconducting materials.~\cite{ceru2-2, daigo2013} Also, low gamma value on LaIr$_{3}$ may be due to the absence of \textit{f}-orbitals near Fermi level and further details can be found on band structure figures later on this manuscript.

Strength of the electron-phonon coupling is estimated with $\frac{\Delta C}{\gamma T_{c}}$.~\cite{padamsee1973} Which is calculated as 1.22 (specific heat jump $\frac{\Delta C}{T_{c}}$ = 14 $\frac{mJ}{mol~K^{2}}$). Calculated value is lower than 1.43 for a conventional BCS superconductor and it suggests that LaIr$_{3}$ is a weakly electron$-$phonon coupled superconductor.
Debye temperature $\Theta _{D}$ is calculated by using $\beta = nN_{A}\frac{12}{5}\pi ^{4}R\Theta _{D}^{-3}$, where $R = 8.314~\frac{J}{mol~K}$, $\textit{n}$ is the number of atoms per formula unit, and $N_{A}$ is Avogadro\textquoteright s number.
The calculated Debye temperature for LaIr$_{3}$ is 366 K, which is slightly higher than that of CaIr$_{2}$.~\cite{neel_cair2}
McMillan formula
$\lambda _{ep} = \frac{1.04 + \mu ^{*} ln\frac{\Theta _{D}}{1.45T_{c}}}{(1-0.62\mu ^{*}) ln\frac{\Theta _{D}}{1.45T_{c}}-1.04}$ is generally used to calculate the strength of the electron-phonon coupling.~\cite{mcmillan1968, poole1999} 
Dimensionless electron-phonon coupling constant $\lambda _{ep}$ is within the 
McMillan\textquoteright s model, which, in Eliashberg theory, is related to the phonon spectrum and the density of states. 
Attractive interaction represents by the parameter $\lambda _{ep}$, while the screened Coulomb repulsion represents by the second parameter $\mu ^{*}$.
The electron-phonon coupling constant ($\lambda _{ep}$) for LaIr$_{3}$ is calculated as 0.57 by using the Debye temperature $\Theta _{D}$, critical temperature $T_{c}$, and making the common assumption that $\mu ^{*} = 0.15$,~\cite{mcmillan1968} 
Weak electron-phonon coupling behavior from $\lambda _{ep}$ = 0.57 agrees well with $\frac{\Delta C}{\gamma T_{c}} = 1.22 $.

\begin{figure}[t]
  \centerline{\includegraphics[width=0.5\textwidth]{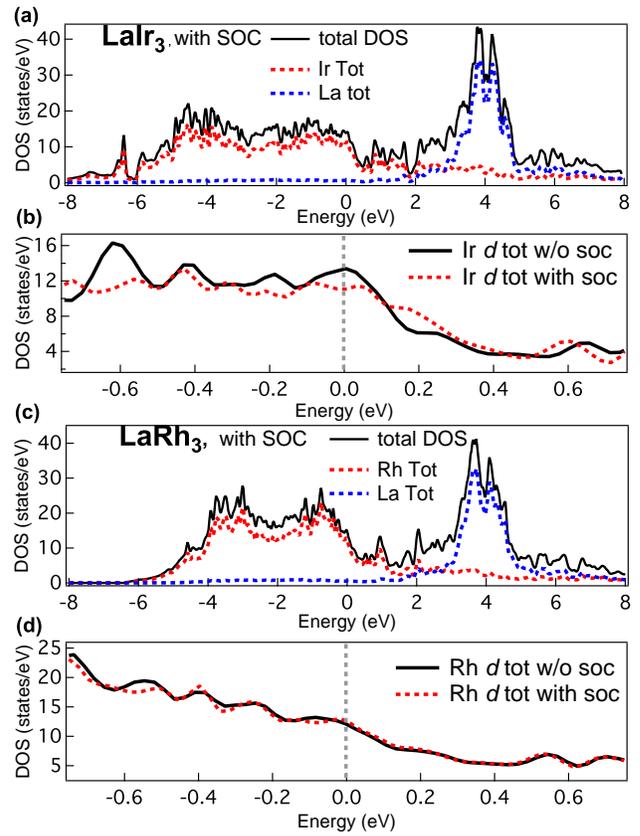}}
  \caption
    {
      (Color online) The density of states of LaIr$_{3}$ and LaRh$_{3}$. (a) The total DOS of LaIr$_{3}$, as well as the contributions of La and Ir states respectively. Spin-orbit coupling (SOC) is included. (b) Ir \textit{d}-band DOS with and with-out SOC. (c) The total DOS of LaRh$_{3}$, as well as the contributions of La and Rh states respectively. Spin-orbit coupling (SOC) is included. (d) Rh \textit{d}-band DOS with and with-out SOC.    
    }
  \label{Fig4}
\end{figure}

\begin{figure}[t]
  \centerline{\includegraphics[width=0.5\textwidth]{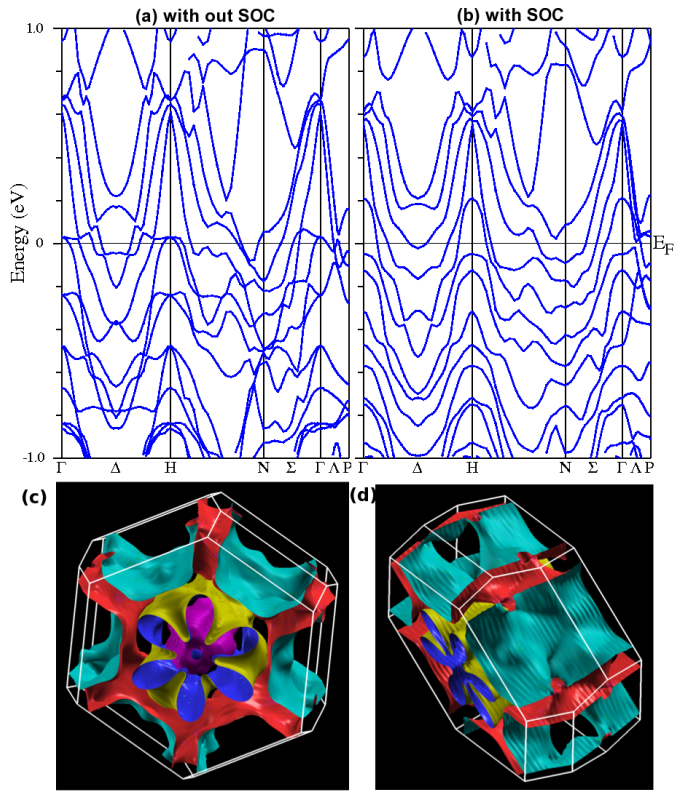}}
  \caption
    {
      (Color online) Electronic band structure of LaIr$_{3}$. (a) Band-structure with-out SOC and (b) band-structure with SOC. Complex Fermi surface shows on (c) and (d) for merged bands.  
    }
  \label{Fig5}
\end{figure}

The measured magnetic susceptibility of LaIr$_{3}$ at 10 K is $\chi = 1.44 \times 10^{-6} \frac{emu~m^{3}}{mol~f.u.}$, which can be considered as a spin susceptibility. This allows for an estimate of the Wilson ratio $R_{W} = \frac{\pi ^{2} k_{B}^{2} \chi _{spin}}{3\mu _{B}^{2}\gamma } = 0.21$, which is smaller than that of the free electron value of 1. Also, the coefficient of the quadratic resistivity term can be normalized with the effective mass term from the heat capacity, which results in the Kadowaki-Woods ratio $\frac{A}{\gamma ^{2}}$. This ratio is found to be 0.39~$a_{0}$, where $a_{0}$ = 10$^{-5} \frac{m \Omega ~cm}{(mJ / mol~K)^{2}}$. $R_{W}$ and $\frac{A}{\gamma ^{2}}$ both indicate that LaIr$_{3}$ is a weakly-correlated electron system.~\cite{jacko2009, wilson1975, yamada1975}

All the measured and calculated superconducting parameters of LaIr$_{3}$ are summarized and presented in Table.~\ref{tab:1}.

Fig.~\ref{Fig4} shows the analysis of the density of states (DOS) of isostructural LaIr$_{3}$ and LaRh$_{3}$.
It can be clearly observed that the Ir states (Fig.~\ref{Fig4}(a)) and Rh states (Fig.~\ref{Fig4}(c)) dominate near the Fermi level, respectively. The contribution of La states is almost negligible at the Fermi level in both systems. 
Detailed analysis of the DOS calculated with and without SOC shows that it is affecting the LaIr$_{3}$ system significantly, but not LaRh$_{3}$.
To further investigate the effect of SOC, the partial DOS of both systems is plotted in Fig.~\ref{Fig4}(b) and (d). It can be clearly observed that the Ir-\textit{d} states and Rh-\textit{d} states are dominant near the Fermi level. However, the Ir compound clearly shows a stronger SOC effect, while there is no such effect in the Rh system. This is similar to other Ir based superconducting systems such as CaIr$_{2}$.~\cite{neel_cair2}
The partial DOS analysis shows that the total DOS is dominated by the contributions from the Ir sublattice, and that the contribution from the La atoms near the Fermi level is almost negligible.
Given the radical change in the Fermi surface (see Fig.~\ref{Fig5}) due to the spin orbit coupling compared to the hypothetical case where no spin orbit coupling is present, one can speculate that SOC has an effect on the superconducting properties of LaIr$_3$. This is further strengthened by the fact that T$_{c}$ is lower in the isostructural Rh analog, where SOC is negligible.~\cite{larh3}
LaIr$_3$ is thus a rare-example of a lanthanide superconductor that is superconducting due the effect of 5\textit{d} electrons. 

Fig.~\ref{Fig5} shows the calculated band structure of LaIr$_{3}$. The SOC effect can be clearly observed near the Fermi level of Fig.~\ref{Fig5}(b) and significant differences can be observed between the Fig.~\ref{Fig5}(a) without SOC and Fig.~\ref{Fig5}(b) with SOC. Cubic laves phase superconductor CaIr$_{2}$~\cite{neel_cair2} also shows the similar behavior. 
3D metallic nature is observed from the band structure of LaIr$_{3}$; several bands with large dispersion cross the Fermi level. 
Ir \textit{5d}-orbitals are the only bands visible near Fermi level. This suggests that the superconducting electrons are strongly effected by the Ir sub-lattice of LaIr$_{3}$.
Calculated complex Fermi surface (See Fig.~\ref{Fig5}(c) and (d)) is a result of many bands crossing the Fermi energy. 

LaIr$_{3}$ is an rare-earth based superconductor with the presence of \textit{f}-orbitals. However, this study suggests that superconductivity of this system is derived by 5\textit{d}-orbitals of Ir-sublattice and there is no any effect from La \textit{f}-orbitals. And also, this study further suggests that strong effect of SOC on Ir-5\textit{d} bands, which is very commonly found on many other Ir based superconductors.~\cite{ceirin5, daigo2013, ca3irge32014, cairge32010, daigo2014, neel_cair2}
To investigate further the role of rare-earth elements and SOC effect from Ir 5\textit{d} bands, other rhombohedral structure type \textit{R}Ir$_{3}$ materials with \textit{R}-lanthanides (with 4\textit{f})~\cite{lair3_2} will be an excellent playground. Specially materials with partially filled 4\textit{f} orbitals, such as Pr, Ce, \textit{etc.,} will show magnetism effect and possible superconductivity. 
Also, it may be very interesting and important to see the real band structure of these \textit{R}Ir$_{3}$ materials by using angle resolved photoelectron spectroscopy (ARPES).
And specially CeIr$_{3}$~\cite{lair3_2} is reported to be a superconductor but there is no detail studies reported so far. We think that CeIr$_{3}$ and doped series of LaIr$_{3}$ to CeIr$_{3}$ will be an excellent study to further investigate, which will possibly show unusual physical properties due to the effect of magnetic rare earth element of Ce with \textit{f} orbitals and possible strong correlations effects of Ce 4\textit{f} and Ir 5\textit{d} bands.  

\section{Conclusion}

Here, we have reported the synthesis and characterization of LaIr$_{3}$, which displays superconductivity below $T_{c}=3.3 K$.  LaIr$_3$ is a superconductor where the bands near the Fermi surface are dominated by Ir 5\textit{d} states that are strongly affected by SOC.  Thus, it is one of the few examples of a lanthanide-based superconductor where 5\textit{d} electrons play the dominate role in the superconductivity. The superconducting parameters obtained from physical properties measurements of LaIr$_3$ suggest that it is a weakly-coupled BCS-type superconductor.  These results motivate future work focused on exploring intermetallic Ir-based superconductors that form in noncentrosymmetric structure types.  Strong SOC in these systems would then be antisymmetric, which often leads to exotic and unconventional superconductivity.

\section{acknowledgments}

DPY acknowledges support from the NSF through Grant No. DMR-1306392.

\end{document}